\documentclass[runningheads]{cl2emult}

\usepackage{makeidx}  
\usepackage{graphicx} 
\usepackage{subeqnar} 
\usepackage{multicol} 
\usepackage{cropmark} 
\usepackage{eso}      
\makeindex            

\begin{document}
%
%
%
\def\spose#1{\hbox to 0pt{#1\hss}}
\def\lta{\mathrel{\spose{\lower 3pt\hbox{$\sim$}}
    \raise 2.0pt\hbox{$<$}}}
\def\gta{\mathrel{\spose{\lower 3pt\hbox{$\sim$}}
    \raise 2.0pt\hbox{$>$}}}
%
%
\def\arcsec{\hbox{$^{\prime\prime}$}}
\def\farcm{\hbox{$.\mkern-4mu^\prime$}}
\def\farcs{\hbox{$.\!\!^{\prime\prime}$}}
%
%
\title*{Evidence for Massive Black Holes\protect\newline 
        in Nearby Galactic Nuclei} 
\toctitle{Evidence for Massive Black Holes
\protect\newline in Nearby Galactic Nuclei} 
%
%
\titlerunning{Black Holes in Galactic Nuclei}

\author{Tim de Zeeuw\inst{\null}}
\authorrunning{Tim de Zeeuw}
\institute{Leiden Observatory, Postbus 9513, 2300 RA Leiden, The Netherlands}

\maketitle              

\begin{abstract}
Masses of black holes in nearby galactic nuclei can be measured in a
variety of ways, using stellar and gaseous kinematics. Reliable black
hole masses are known for several dozen objects, so that demographic
questions can start to be addressed with some confidence. Prospects
for the near future are discussed briefly.
\end{abstract}

\section{Introduction}
\label{sec:intro}
Active galaxies and quasars are powered by physical processes in an
accretion disk surrounding a massive black hole \cite{lb69,ree84}.
The observed number of active galaxies increases towards high redshift
$z$, so that many `normal' galaxies must have been active in the past
\cite{sol82}. This implies that inactive massive central black holes
must lurk in the nuclei of nearby normal galaxies. In the past decade,
much work has been done to measure the masses of these black holes, to
establish the relation between black hole mass and the global/nuclear
properties of the host galaxy, and to understand the role these
objects play in driving internal dynamical evolution
\cite{mer99,zee96}.

A black hole of mass $M_{\rm BH}$ in a galactic nucleus dominates the
gravitational potential inside the so-called radius of influence which
is usually defined as $r_{\rm BH}= GM_{\rm BH}/\sigma^2$, where $G$ is
the gravitational constant, and $\sigma$ is the characteristic
velocity dispersion in the host galaxy. In physical units
\begin{equation}\label{eq:one}
r_{\rm BH} \sim 0.4 \Bigl( {M_{\rm BH} \over 10^6 M_\odot} \Bigr) 
                \Bigl( {100 \, \hbox{\rm km/s} \over \sigma} \Bigr)^2 
                \, \hbox{\rm pc}. 
\end{equation}
For a galaxy at distance $D$, $r_{\rm BH}$ corresponds to an angular
size
\begin{equation}\label{eq:two}
\theta_{\rm BH} \sim 
          0\farcs1 \Bigl( {M_{\rm BH} \over 10^6 M_\odot} \Bigr) 
                \Bigl( {100 \, \hbox{\rm km/s} \over \sigma} \Bigr)^2  
                \Bigl( {1 \, \hbox{\rm Mpc} \over D} \Bigr). 
\end{equation}
Inside the sphere of influence, the black hole generates a central
cusp in the density distribution of the stars. The resulting density
and luminosity $\propto r^{-\gamma}$ with $3/2 \leq \gamma \leq 9/4$
\cite{qhs95}. The typical velocities scale $\propto r^{-1/2}$ with
radius.

While many early-type galaxies have cusps in their central luminosity
profiles \cite{car+97,fab+97,jaf+94}, these by themselves are not
proof of the presence of a black hole, as other processes can generate
such cusps \cite{kor93}.  For this reason most studies concentrate on
obtaining spectroscopy at the smallest angular scales, to find
evidence for the expected Keplerian rise in the velocities of stars
and gas. This requires high spatial resolution. For example, the $3
\times 10^9 M_\odot$ black hole in the nucleus of the galaxy M87
\cite{har+94} in the Virgo cluster has $\theta_{\rm BH} \approx
1\farcs5$.  The general approach is to determine the {\it luminous}
mass in stars from the observed surface brightness distribution, and
to compare this with the {\it dynamical} mass derived from kinematic
measurements of stars or gas \cite{ho98,ric+99}. If one can show that
the mass density inside a certain radius is larger than anything that
can be produced by normal dynamical processes, then the object is
considered to be a black hole. In a few cases it is possible to find
direct evidence for the presence of a relativistic object
(\S\ref{sec:gas}).

\begin{figure}[t]
\centering
\includegraphics[width=.64\textwidth,angle=-90.]{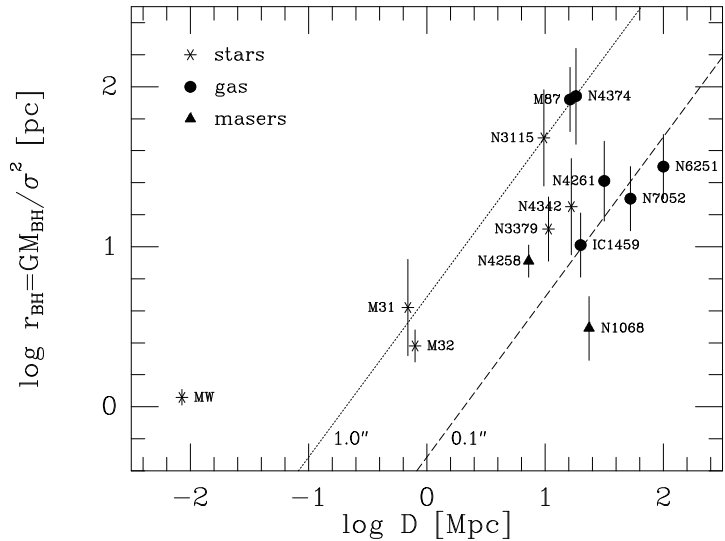}
\caption[]{The radius of influence $r_{\rm BH}= GM_{\rm BH}/ \sigma^2$
of central black holes with well-determined masses versus the distance
of the host galaxy. The masses were derived from stellar kinematics
(stars), gas kinematics (dots) and VLBI measurements of masers
(triangles). The solid line corresponds to an angular size
$\theta_{\rm BH} = 1\farcs0$, and the dotted line corresponds to
$0\farcs1$. The error bars represent the quoted uncertainty in $M_{\rm
BH}$ \cite{bac00,bow+98,cb99,eg97,edb99,ffj96,ff99,geb+00a,ghe+98,gre+96,mac+97,mb98,mzr+97,miy+95,vk+00}. }
\label{fig:rix}
\end{figure}

Figure~\ref{fig:rix} shows $r_{\rm BH}$ as defined in eq.\
(\ref{eq:one}) for the best published black hole mass determinations
versus the distance to the host galaxy. Lines of constant angular
resolution $\theta_{\rm res}$ run diagonal. The early determinations
clustered near $\theta_{\rm res}\approx 1\farcs0$ \cite{rix93}, but
HST has pushed this to $\theta_{\rm res}\approx 0\farcs1$. Detailed
modeling has shown that, depending on the internal dynamical structure
of the host galaxy, the effects of the black hole often are visible
only inside projected radii that are significantly smaller than
$r_{\rm BH}$ (e.g., eq.\ (4.2) in \cite{qia+95}). This suggests that
measured masses corresponding to $\theta_{\rm BH} \sim \theta_{\rm
res}$ should be treated with caution, as they are likely to be
overestimates (see \S\ref{sec:demo}). At present, only VLBI
measurements can probe the regime $\theta_{\rm res} < 0\farcs1$.

\vfill\eject
\section{Stellar dynamical modeling}
\label{sec:stars}
The nearest galactic nucleus is the center of our own Galaxy.  Despite
the large foreground extinction, it is possible to resolve individual
stars in the Galactic center in the infrared, and to measure not only
their radial velocities, but also their proper motions
\cite{eg97,ghe+98}, and accelerations \cite{ghe00}!  Dynamical
modeling of this remarkable data provides unequivocal proof that our
Galaxy contains a central black hole of nearly three million solar
masses.

The dynamics of the nuclei of nearby early-type galaxies can be probed
with stellar absorption-line spectroscopy of the integrated
light. This generally requires long exposure times. The orbital
structure in these systems is rich \cite{mer99,zee96}, so a true
inward increase of the mass-to-light ratio $M/L$ must be distinguished
from radial variation of the velocity anisotropy. This can be done by
measuring the shape of the line-of-sight velocity distribution
\cite{ger93,mf93}. The orbital structure is related to the intrinsic
shape of the galaxy. This can be constrained by measurements along
multiple position angles \cite{bs00,mcz+98}, or, even better, by
integral-field spectroscopy (\S\ref{sec:next}).

Determination of the black hole mass and the orbital structure
requires construction of dynamical models. There has been a steady
increase in the sophistication of model construction in the past
decade. Early isotropic spherical models were replaced by anisotropic
spheres, and then by axisymmetric models with a special orbital
structure (phase-space distribution function $f=f(E, L_z)$ where $E$
is the orbital energy and $L_z$ is the angular momentum component
parallel to the symmetry axis, e.g., \cite{qia+95}). More recently
axisymmetric models with the full range of possible anisotropies, and
multiple components have been used. The first such study was done for
M32, and included ground-based data along four position angles and
eight {\tt FOS} pointings \cite{mcz+98}. The model was constructed by
a version of Schwarzschild's \cite{sch79} numerical
orbit-superposition method, which fits the surface brightness
distribution as well as all kinematic observables \cite{cre+99}.

\begin{figure}[t]
\centering
\includegraphics[width=.6\textwidth,clip=]{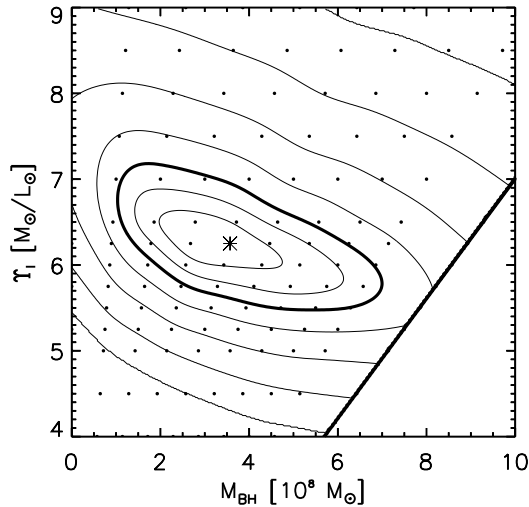}
\caption[]{The black hole in NGC~4342. Contours indicate the
goodness-of-fit $\chi^2$ to the observed photometry and kinematics as
a function of black hole mass $M_{\rm BH}$ and $\Upsilon_I$, the
$I$-band stellar mass-to-light ratio. Dots indicate dynamical models
that were constructed. The asterisk denotes the best fit.  The first
three contours surrounding it are the 68.3\%, 95.4\% and 99.7\%
confidence levels, while the subsequent contours correspond to a
factor of two increase in $\chi^2$ \cite{cb99}. }
\label{fig:n4342}
\end{figure}

Another example of this approach is provided by the E7/S0 galaxy NGC
4342 in Virgo. This is a low-luminosity object, seen nearly edge-on,
with a prominent nuclear stellar disk.  Cretton \& van den Bosch
\cite{cb99} used ground-based major-axis kinematics from the WHT and
five {\tt FOS} pointings \cite{bjm98}, and compared these with 
general axisymmetric models containing a spheroid, a stellar disk, a
nuclear disk and a black hole. Free parameters in the modeling were
$M_{\rm BH}$ and the stellar mass-to-light ratio $\Upsilon$. Contours
of constant $\chi^2$ in the $(\Upsilon, M_{\rm BH})$-plane show that
the best fit is obtained for $M_{\rm BH} \sim 3\times 10^8 M_\odot$,
but with a significant uncertainty (Figure~\ref{fig:n4342}).

To date, black hole masses have been derived in this way for about
fifteen objects. The full analysis has been published only for M32
\cite{mcz+98}, N4342 \cite{cb99}, and NGC3379 \cite{geb+00a}, all
based on ground-based and {\tt FOS} data.  Black hole masses based on
three-integral axisymmetric modeling of {\tt STIS} absorption-line
spectroscopy have been reported for a dozen objects by Gebhardt's
group \cite{geb+00b}, but the data and the models have not yet been
published.

While in some cases axisymmetric models are consistent with the data,
in others they clearly are inadequate (\S\ref{sec:next}).  The nucleus
of M31 has been known to be asymmetric since 1974
\cite{lds74}. Observations with the integral-field spectrograph {\tt
TIGER} revealed that the black hole resides in the secondary peak in
the brightness distribution \cite{bac+94}, and led to models with an
asymmetric distribution of stars \cite{tre95}. Measurements with {\tt
OASIS} on the CFHT \cite{bac00} have shown that HST spectroscopy with
{\tt FOC} \cite{sta+99}, {\tt FOS} and {\tt STIS} along the apparent
symmetry axis of the eccentric structure has in fact missed the
kinematic major axis. A comprehensive non-axisymmetric model that fits
this data will teach us much about the structure and formation of this
nearby nucleus.\looseness=-2

Some nearby galactic nuclei are shrouded in dust, and their internal
kinematics is best probed at longer wavelengths.  This is now possible
through the availability of near-IR spectrographs, which employ the CO
bandhead at 2.3$\mu$ to derive the stellar kinematics. Anders
\cite{and99} used the MPE-built {\tt 3D} integral-field unit to show
that the derived nuclear kinematics in the largely dust-free nearby
galaxy NGC 3115 agrees with the kinematics measured at shorter
wavelengths \cite{edb99}, demonstrating that this approach works. {\tt
3D} observations of the luminous merger remnant NGC 1316 (Fornax A)
confirmed the central $\sigma$ of $\approx$230 km/s obtained at short
wavelengths from the ground and with {\tt FOS} \cite{sha99},
suggesting $M_{\rm BH} \lta 10^8 M_\odot$ \cite{dav00}.  Instruments
such as {\tt SINFONI} will probe nearby dusty nuclei with a resolution
similar to HST \cite{men+00}.

\section{Gas kinematics}
\label{sec:gas}

{\it Optical emission lines.} The nuclei of active early-type
galaxies, and those of most spirals, contain extended optical
emission-line gas. Its kinematics can be used to constrain the central
mass distribution. In this case the exposure times can be relatively
short, but care must be taken to model possible non-circular motions
and the effects of turbulence in the gas disks.

High spatial resolution emission-line kinematics ({\tt FOC}, {\tt FOS}
and {\tt STIS}), together with careful modeling, has been published
for six cases: M87, NGC 4261, NGC 4374, NGC 6251, NGC 7052, IC 1459
\cite{bow+98,ffj96,ff99,mac+97,mb98,vk+00}. These very luminous
early-type galaxies cover a modest range in total luminosity, but
the derived black hole masses vary by a factor of ten.

\begin{figure}[b]
\centering
\includegraphics[width=1.0\textwidth]{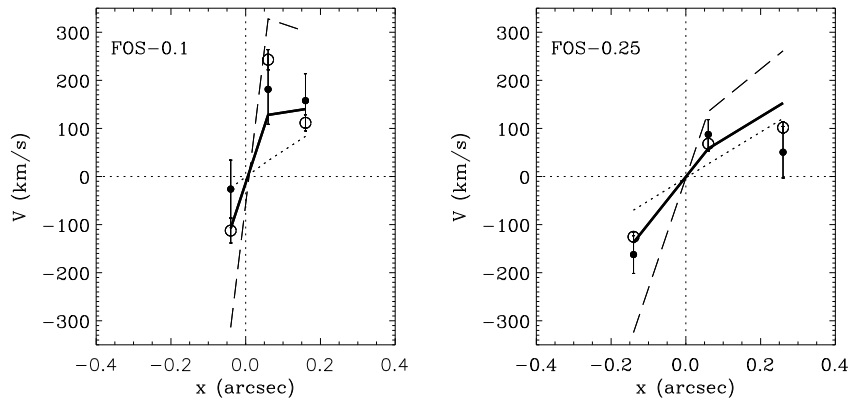}
\caption[]{Emission-line gas kinematics of the nucleus of IC1459,
derived from three {\tt FOS} pointings with the $0\farcs1$ (left) and
three with the $0\farcs25$ apertures (right). Rotation velocities $V$
were derived from H$\alpha$$+$[NII] (open circles) and H$\beta$
measurements (dots). The heavy solid line is the prediction of a model
with $M_{\rm BH}=1.0\times 10^8 M_\odot$. Dotted and dashed curves are
for $M_{\rm BH} = 0$ and $7\times 10^8 M_\odot$, respectively
\cite{vk+00}.  }
\label{fig:i1459}
\end{figure}

An example of this approach is provided by the E3 galaxy IC~1459.
This giant elliptical hosts a compact nuclear radio source, has a
counter-rotating stellar core ($\sim 10''$), a shallow cusp in the
luminosity profile, and a blue central point source
\cite{car+97,fi88}.  {\tt FOS} kinematics of the emission-line gas
associated with the nuclear dust-lane reveals a disk in rapid rotation
(Figure~\ref{fig:i1459}), with significant velocity dispersion.
Detailed models for the gas motions which include the effects of
turbulence show that $M_{\rm BH} \approx 2.5 \times 10^8 M_\odot$
\cite{vk+00}, which is a factor of ten smaller than the mass suggested
by axisymmetric $f(E, L_z)$ models of the ground-based stellar
kinematics along the major axis. A general stellar dynamical model
that incorporates major-axis {\tt STIS} spectroscopy and ground-based
spectroscopy along four position angles is under
construction.\looseness=-2

Kinematics of emission-line gas will provide many black hole masses in
the near future. Ongoing HST programs include H$\alpha$ emission-line
spectroscopy with {\tt STIS} for 21 radio-loud ellipticals (PI Baum)
and 54 Sb/Sc spirals (PI Axon). These studies generally use three
parallel slits to measure possible deviations from simple circular
motion.

\smallskip
\noindent
{\it Masers}. Nuclear maser emission can be measured with VLBI
techniques in a few spiral galaxies. This achieves the highest spatial
resolution to date, but is possible only when the circumnuclear disk
is nearly edge-on. A search of $\approx$700 nuclei revealed maser
emission in 22 cases, and disk-like kinematics in six of these
\cite{mor00}. The best black hole mass determinations are for NGC
1068, NGC 4258, and NGC 4945 \cite{gre+96,miy+95}.

\smallskip
\noindent
{\it X rays}. Recently, it has become possible to measure the profile
of the Fe K$\alpha$ line at 6.4 keV in the nuclei of nearby Seyfert
galaxies \cite{nan+97}.  The width of this line approaches
$\approx$$10^5$ km/s, which is direct evidence that the emitting gas
must be near the Schwarzschild radius of a relativistic object. The
mass of the black hole, and possibly its spin, can be derived from the
detailed shape of the line profile \cite{fab+89,mkm00}. The latest
generation of X-ray telescopes, in particular CHANDRA and XMM, will
extend this work to more objects.

\smallskip
\noindent
{\it Reverberation mapping}. The observed time-variation of broad
H$\beta$ emission lines from Seyfert nuclei can be used to derive the
radius of the broad-line region by means of the light travel-time
argument \cite{bk82}. In combination with simple kinematic models for
the motion of the broad-line clouds, this provides an estimate of the
mass of the central object responsible for these motions.  Early mass
determinations appeared to be systematically lower than those obtained
by other means \cite{ho98}, but this discrepancy has now disappeared
\cite{geb+00c}.

\section{Demographics}
\label{sec:demo}
For the past five years, our understanding of black hole demography was
summarized in a diagram of black hole mass $M_{\rm BH}$ versus
absolute bulge luminosity $L_{\rm bulge}$, where the `bulge' was taken
to be the entire galaxy for an elliptical or lenticular, or the actual
bulge in case of a spiral \cite{kr95,ric+99}. Early work based on
simple axisymmetric Jeans models and ground-based kinematics
\cite{mag+98} was interpreted as evidence for a tight correlation
of $M_{\rm BH}$ with $L_{\rm bulge}$, and hence with the total
mass of the bulge.  Much effort was spent in trying to reproduce this
correlation in models of galaxy formation, and in relating it to the
energy production in quasars \cite{ric+99}.   

\begin{figure}[t]
\centering
\includegraphics[width=0.47\textwidth,angle=-90.]{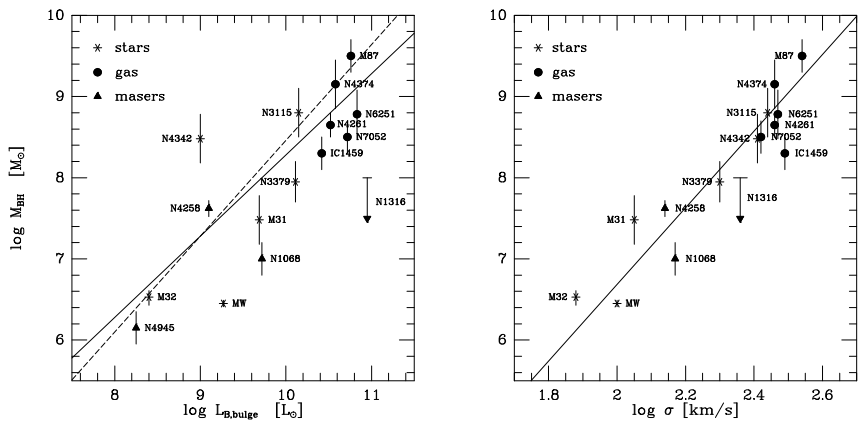}
\caption[]{Demography of black holes. Left: $M_{\rm BH}$ versus total
absolute luminosity $L_{\rm B,bulge}$ of the host bulge or
spheroid. Coding of points is as in Fig.~\ref{fig:rix}. The dashed
line is the correlation based on $f(E, L_z)$ models \cite{mag+98}. The
solid line is the prediction for adiabatic growth models
\cite{vdm99}. Recently published measurements generally give lower
masses, indicating mild radial anisotropy. They display a significant
scatter at fixed $L_{\rm B,bulge}$. There is an observational bias
against detecting small black hole masses in luminous galaxies. Right:
$M_{\rm BH}$ versus velocity dispersion $\sigma$ of the host bulge or
spheroid.  The solid line is the relation proposed in \cite{mf00}. }
\label{fig:mbh-sigma}
\end{figure}

More appropriate dynamical models combined with HST kinematics have
generally resulted in a downward revision of the earlier masses
\cite{geb+00b,ho98,vdm99}. Figure~\ref{fig:mbh-sigma}{\it a} is the
resulting demography diagram for the best determinations.  It shows a
rough correlation, but with a large range in $M_{\rm BH}$ at fixed
$L_B$.  The masses are consistent with various black hole formation
scenarios \cite{vdm99,sti98}, and they agree with the quasar light
prediction for reasonable efficiencies \cite{ric+99,sol82}. However,
there is an observational bias against detecting small black holes in
large galaxies, i.e., for cases where the radius of influence is
simply smaller than the spatial resolution currently achievable.

Absolute galaxy luminosity correlates with velocity dispersion
\cite{fj76}, so $M_{\rm BH}$ is expected to correlate with $\sigma$,
measured outside the radius of influence of the black hole.  As
illustrated in Figure~\ref{fig:mbh-sigma}{\it b}, the scatter in this
correlation is considerably less than in the $M_{\rm BH}$ versus
$L_{\rm bulge}$ correlation \cite{fm00,geb+00b}. A good fit is
provided by $M_{\rm BH} = 1.30(\pm0.36)\times 10^8 M_\odot (\sigma/
200 \, \hbox{km~s}^{-1} )^{4.72(\pm0.36)}$, where $\sigma$ is the
velocity dispersion at $R_e/8$, and $R_e$ is the effective radius
\cite{mf00}. This suggests yet another link between global and nuclear
galaxy properties, but it is not understood what causes this link, or
whether there are other parameters involved. Some of the galaxies in
the diagram are not axisymmetric, so the black hole mass derived from
axisymmetric modeling may need revision.

Substitution of the $M_{\rm BH}$ versus $\sigma$ correlation in
eq.\ (\ref{eq:two}) provides the following estimate of the formal radius
of influence of the black hole:
\begin{equation}\label{eq:three}
\theta_{\rm BH} \sim 
          0\farcs4 \Bigl( {\sigma \over 100 \, \hbox{\rm km/s}} \Bigr)^{2.7}  
            \Bigl( {1 \, \hbox{\rm Mpc} \over D} \Bigr). 
\end{equation}
As noted in \S\ref{sec:intro}, the actual radius of influence may be
smaller.  Assuming that eq.\ (\ref{eq:three}) is valid for all
galaxies (which is not proven) allows one to use a ground-based
velocity dispersion measurement to estimate the minimum resolution
needed to measure a reliable black hole mass.

\vfill\eject
\section{Next steps} 
\label{sec:next}
There is strong evidence for the existence of massive black holes in
most galactic nuclei, based on different measurement approaches. Our
understanding of black hole demographics is still limited, because the
number of reliable mass measurements is relatively modest. Significant
improvement is expected through spectroscopic surveys of galactic
nuclei with emission-line gas, and through further work on stellar
absorption-line kinematics.

\begin{figure}[t]
\centering
\includegraphics[width=1.0\textwidth]{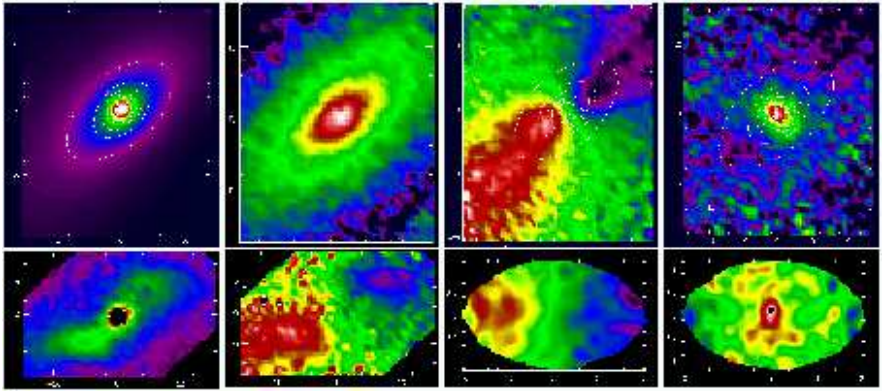}
\caption[]{Integral-field spectroscopy of the E6 galaxy NGC 3377. Top,
from left to right: reconstructed intensity, stellar Mg b index, mean
velocity, and velocity dispersion. Bottom, from left to right: gas
intensity (OIII $\lambda$5007) and velocity, and stellar mean velocity
and velocity dispersion of the nuclear region.  The first six panels
are based on a two-hour exposure with {\tt SAURON}.  The field shown
is $30''\times39''$. The last two panels are based on a 3.5 hour
exposure with {\tt OASIS}, with $0\farcs16$ sampling and a
$4''\times4''$ field.  The stars show a striking rotating disk pattern
with the spin axis misaligned $\approx 10^\circ$ from the photometric
minor axis, indicating the galaxy is triaxial. The gas also reveals
non-axisymmetric structures and motions \cite{cop00,cop+00}. }
\label{fig:n3377}
\end{figure}

To make further progress, high-resolution observations of the nuclei
obtained with {\tt STIS} onboard HST, and with adaptive optics from
the ground (e.g., {\tt OASIS} on the CFHT, {\tt SINFONI} on the VLT),
need to be complemented with the wide-field kinematics of the host
galaxy, in order to measure its intrinsic shape and orbital structure
\cite{bs00,zee96}. For this reason, the dynamics groups of the
universities at Lyon, Leiden, and Durham have built the
special-purpose integral-field spectrograph {\tt SAURON} for the WHT,
with high throughput, and a field of view of $33''\times 41''$ sampled
at $0\farcs94\times 0\farcs94$ \cite{bac+00,zee+00}, and are using it
to observe a representative sample of 80 nearby early-type galaxies.

Figure~\ref{fig:n3377} shows integral-field kinematics for the E6
galaxy NGC~3377 \cite{cop00,cop+00}. The {\tt SAURON} maps of the
stellar and gaseous motions reveal that this galaxy is not
axisymmetric. The minor axis rotation persists in the high-spatial
resolution measurements with {\tt OASIS}. NGC 3377 has a steep-cusped
central luminosity profile, and the expectation based on general
dynamical arguments is that this galaxy should be nearly axisymmetric,
at least in the inner regions \cite{mer99,zee96}.  It will be
interesting to probe this galaxy at even higher spatial resolution, to
establish whether the nucleus is axisymmetric and the derived black
hole mass correct, or whether the non-axisymmetry persists, and our
understanding of the nuclear dynamics needs modification. Either way,
the codes for numerical model construction need to be generalized to
triaxial geometry, and to asymmetric systems such as the nucleus of
M31.

The ongoing systematic programs will reveal the black hole
demographics as a function of Hubble type, radio properties, intrinsic
shapes, and internal dynamics. It will also establish the rate of
occurrence of gaseous and stellar nuclear disks, and of nuclear star
clusters, and the importance of black hole driven secular dynamical
evolution for shaping galaxies.

\smallskip
\noindent
It is a pleasure to acknowledge comments and contributions by Nicolas
Cretton, Roeland van der Marel, Gijs Verdoes Kleijn, and the {\tt
SAURON} team.

\clearpage
\addcontentsline{toc}{section}{Index}
\flushbottom
\printindex

\end{document}